\documentstyle[prb,aps,epsfig]{revtex}
\begin{document}
\twocolumn[\hsize\textwidth\columnwidth\hsize\csname @twocolumnfalse\endcsname

\title{A novel iterative strategy for protein design}
\author{Andrea Rossi, Amos Maritan and Cristian Micheletti}
\address{International School for Advanced Studies (SISSA) and
INFM,
Via Beirut 2-4, 34014 Trieste, Italy\\
The Abdus Salam Centre for Theoretical Physics - Trieste, Italy}
\date{\today}
\maketitle
\begin{abstract}
We propose and discuss a novel strategy for protein design.
The method is based on recent theoretical advancements which showed
the importance to treat carefully the conformational free energy of
designed sequences. In this work we show how computational cost can be
kept to a minimum by encompassing negative design features,
i.e. isolating a small number of structures that compete significantly
with the target one for being occupied at low temperature.
The method is succesfully tested on minimalist protein models and
using a variety of amino acid interaction potentials.
\end{abstract}
\vskip 0.3cm
]

\newpage

\section{Introduction}

Since the late 50's it has been established that the native state of a
protein is entirely and uniquely encoded by its amino acids sequence
\cite{BT91,dieci}. One of the fondamental issues in molecular
biology is understanding the relation between protein sequence and
native structure. Remarkably, this relation is not symmetric: while a
given sequence folds into a single structure, a given structure can be
encoded by several homologous sequences. The problem of predicting
the native state of a sequence is commonly known as "protein
folding". Its solution amounts to minimize the energy of the given
peptidic chain over all possible conformations. The design problem,
i.e. finding the sequence(s) that fold into a desired structure, has
also been given a simple and rigorous formulation. At the
"physiological" temperature $\beta_p^{-1}$, the sequences that
correctly design a given target structure, $\Gamma_t$ (their native
state), maximize the Boltzmann probability \cite{11},

\begin{equation}
\label{Boltzmann} P(s,\Gamma_t,\beta_p)=\exp\Big\{-\beta_p [
H_s(\Gamma_t)-F_s(\beta_p)]\Big\} \ ,
\end{equation}

\noindent where $s=(s_1,s_2,\ldots,s_L)$ represents the amino acid
sequence (amino acid $s_1$ at the first position in $\Gamma, \ldots,
s_L$ at the last position in $\Gamma$) and $H_s$ is the energy of $s$
housed on $\Gamma$. $F_s$ in eq. (\ref{Boltzmann}) is the
conformational free energy of the sequence $s$,

\begin{equation} \label{freeenergy}
F_s(\beta)=-\beta^{-1}\ln\Big\{\sum_\Gamma\ \exp[-\beta
H_s(\Gamma)]\Big\} \ ,
\end{equation}

\noindent where the summation is taken over all possible conformations
the sequence $s$ can assume without violating steric constraints.
Maximizing (\ref{Boltzmann}) poses some serious technical difficulties
since, in principle, it entails an exploration of both sequence and
structure space \cite{11,M98a}. Some simplifications have been used in
the past in order to limit the space of sequences; this is
conveniently done by subdividing amino acids into a limited number,
$q\leq20$, of classes \cite{5,undici}. Some approximation schemes have
also been used to reduce the difficulty of calculating $F_s$
\cite{10,11,12}. Reasonable success has been obtained, for example, by
postulating a suitable functional dependence of $F_s$ on $s$
\cite{M98b,ultimo}. Recently, it has also been argued that, despite
the huge number of conformations, $\Gamma$, the most significant
contribution to (\ref{freeenergy}) comes from the closest competitors
of $\Gamma_t$ \cite{indci}. These are limited in number, since they
are among the ones sharing a subset of native contacts with
$\Gamma_t$.

In this article we propose a new technique apt for designing a given
structure, $\Gamma_t$, using a minimal set of structure for
calculating (\ref{freeenergy}). The technique is simple to implement
and does not require to constrain sequence composition and/or to
search solutions with the lowest energy \cite{7}.
Several exact tests have been implemented in order to assess the performance
of the new method with respect to previously proposed techniques.

\section{Theory: the iterative design scheme}

In order to discuss the design method we introduce a Hamiltonian
function, $H_s(\Gamma)$ depending only on coarse grained degrees of
freedom. A commonly used form is the one in terms of the contact matrix
$\Delta_2(\vec r_i,\vec r_j)$ which is 1 when $|\vec r_i-\vec r_j|<a$
with $a\approx 6-8 \AA$, and 0 otherwise. Other forms which smoothly
interpolate between 0 and 1 are also used in practice. Two amino
acids, $s_i$ and $s_j$, which are in contact contribute to the energy
by an amount $\epsilon_2(s_i,s_j)$, a phenomenological symmetric
matrix (see e.g. refs. [\onlinecite{due,tre,nove,mprep}]). Many
body interations can also be easily included in terms of a generalized
contact maps $\Delta_k(\vec r_{i_1},\ldots,\vec r_{i_k})$ depending
only on relative distances and on extra energy parameters
$\epsilon_k(s_{i_1},\ldots,s_{i_k})$. Thus the energy can be written
as

\begin{equation} \label{hamiltonian} H_s(\Gamma)=\sum_{k\geq
2}\sum_{i_1<i_2<\ldots<i_k} \epsilon_k(s_{i_1},\ldots,s_{i_k})
\Delta_k(\vec r_{i_1},\ldots,\vec r_{i_k}).
\end{equation}

Two structures which have the same values for all the $\Delta_k$'s,
i.e. the same generalised contact map, will be regarded as identical.
This useful coarse-graining procedure neglects the fine structure
fluctuacions (e.g. due to thermal excitations) and, for a sequence
$s$ with native state $\Gamma$, allows to define its folding
temperature, $\beta_F^{-1}$, such that

\begin{equation} \label{solution} P(s,\Gamma,\beta)>1/2 \ ,
\end{equation}

\noindent for all $\beta>\beta_F$. Conversely, all $s$'s satsfying
inequality (\ref{solution}) have their unique ground state in $\Gamma$
and folding temperature greater than $\beta^{-1}$.

Based on this observation the novel strategy for protein design can be
formulated in terms of a scheme similar in spirit the one described in
ref [\onlinecite{mprep}]. The essence of the procedure relies on the fact
that the sum in (\ref{freeenergy}) is carried out only on a limited
set of structures $D$. Initially, $D$ contains only the target
structure itself and another structure with a different contact map
and similar degree of compactness (chosen at random or with other
criteria). Upon iterating the procedure several design solutions will
be identified and stored in set $S$. This set is, of course,
initially empty. The steps to be iterated are as follows,

\begin{enumerate}

\item An optimization procedure, like simulated annealing, is
used to explore sequence space and isolate the sequence $\bar
s$ (not already included in $S$), such that

\begin{equation} \label{solution2} \beta [ H_{\bar
s}(\Gamma_t) -\tilde{{\cal F}}_{\bar s}(\beta)]<\ln 2\
. \end{equation}

$\tilde{{\cal F}}$ is calculated approximately by restricting
the sum in (\ref{freeenergy}) over the competitive structures
held in $D$:

\begin{equation} \label{freeappr} \tilde{{\cal
F}}_s(\beta)=-\beta^{-1}\ln \{\sum_{\Gamma\in D}
\exp[-\beta H_s(\Gamma)]\}\ . \end{equation}

\item Then the lowest energy state(s), $\bar \Gamma$ of $\bar s$ are
identified and the corresponding energy compared with that obtained by
$\bar s$ on $\Gamma_t$. By definition, if $\bar{\Gamma} \not=
\Gamma_t$ and $H(\bar s, \bar \Gamma)\leq H(\bar s, \Gamma_t)$, then
$\bar{s}$ is not a solution to the design problem and $\bar \Gamma$ is
added to $D$. Otherwise, $\bar s$, is added to the set of known
solutions, $S$.
\end{enumerate}

The iterative procedure is repeated from step 1. The scheme stops when
it is impossible to find sequences, satisying (\ref{solution2})
not already included in $S$, or when a sufficient number of solutions
has been retrieved. It is easy to see, using (\ref{solution2}) and
(\ref{freeappr}) that, in step 2, it can never happen that a newly
chosen $\bar{\Gamma} \not= \Gamma_t$ is already contained in $D$.
Thus, at each iteration, new informations are collected, either in the
form of a putative solution (added to $S$) or as a new decoy (added to
$D$).

Notice that, if the exact form of $F_s$ were used instead of
(\ref{freeappr}), then the sequences in $S$ would have a folding
temperature greater than $\beta^{-1}$. However, since approximation
(\ref{freeappr}) leads to systematically overestimating $F_s(\beta)$,
it is not guaranteed that the selected sequences have $\beta_F^{-1} <
\beta^{-1}$. The inequality should however be satisfied to a better
extent for larger decoys sets.

The method outlined here is rigorous and its iterative application
allows, in principle, to extract all sequences designing a given
structure. Its pratical implementation may encounter difficulties
at step 2, where it is required to find the low energy conformation(s)
of a sequence. Sequences selected at step 2 with a low $\beta$ will
correspondingly have a high folding temperature and are expected to be
good folders. Hence, it is plausible that identifying the
corresponding low energy states is much simpler than solving the
general folding problem.  In fact, we have gathered numerical evidence
showing that strategy can be stopped as soon as a one finds a
structure where is attained an energy lower than on $\Gamma_t$ (even
if true native state has still lower energy). Notice that it is still
necessary to have folding technique to allow to test if the design
procedure is successful. Our iterative scheme is able to use
informations of failed attempts in order to improve design at
subsequent iterations.

\section{Methods: implementation and test of the procedure}

To carry out a rigorous and exhaustive test of the proposed strategy
we have restricted the space of structures by discretizing the
positions of amino acids, $\vec r_i$. We choose to follow the common
practice of restricting the $\vec r_i$'s to occupy the nodes of a
cubic lattice \cite{undici,18,8,M98a,steric}. This semplification
allows for an exhaustive search of the whole conformation space for
chain of a few dozen residues, albeit at the expenses of an accurate
representation of protein structures, as discussed in
ref. [\onlinecite{PL95}]. To mimic the high degree of compactness
found in naturally occurring proteins, we first considered all the
maximally compact self-avoiding walks of length $L=27$ embedded in a
$3\times3\times3$ cube. There are 103346 distinct oriented walks
modulo rotations and reflections. This restriction is a good
approximation if the interaction energies between amino acids are
sufficiently negative, so that compact conformations are favoured over
loose ones. Without loss of generality we adopt a Hamiltonian where
only pairwise interactions are considered (corresponding to $k=2$ in
(\ref{hamiltonian})). If the interaction energies are sufficiently
attractive it is guaranteed that the native state is compact. Step 2
of the iterative procedure was carried out in two distinct ways. In a
first attempt we found the true lowest energy state of $s$ by
exhaustive search. In a second attempt we tried to mimic the
difficulty of finding the ground state in a realistic context and
hence carried out a random partial exploration of the structure
space. Although the first method was expected to be more efficient
than the second, their performance turned out to be almost identical,
as we discuss below.

\begin{figure}
\begin{center}
\epsfig{figure=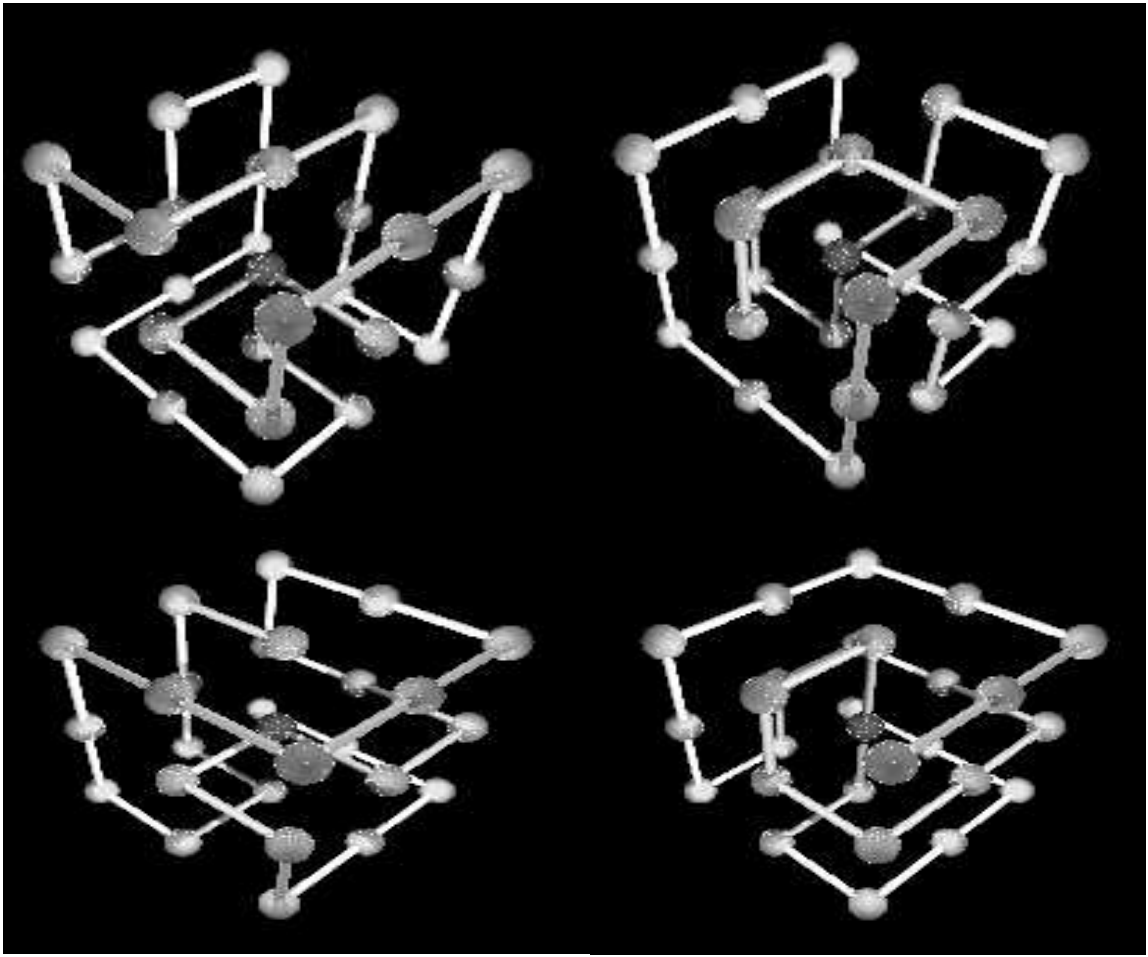,width=6.0in,height=5.0in}
\end{center}
\caption{The target structures $\Gamma_1$ (top left), $\Gamma_2$
(top right),$\Gamma_3$(bottom left), $\Gamma_4$ (bottom right).}
\label{fig:structs}
\end{figure}

The four target conformations used to test the procedure are given in
Table \ref{structures} and shown in Fig \ref{fig:structs}a-d. We used
three possible choices for the $\epsilon$'s. First, we adopted the
standard 2-class HP model with $\epsilon_{\rm HH}=-1-\alpha$ and
$\epsilon_{\rm HP}=\epsilon_{\rm PP}=-\alpha$. $\alpha$ is a suitable
constant ensuring that native conformations are compact. Since all
conformations considered here have the same number of contacts the
value of $\alpha$ is irrelevant and will be omitted from now on. The
second case is a 6-class model and the $\epsilon$'s are shown in Table
\ref{tab:6-class}. For the last case we considered the full repertoire
of 20 amino acids used the Miyazawa and Jernigan energy parameters
given in Table 3 of ref. [\onlinecite{due}]. With the standard HP
parameters, structures $\Gamma_1-\Gamma_4$ have various degree of
designability. The latter is defined as the number of sequences
admitting them as unique ground states \cite{wing} . Hence, the
encodability of $\Gamma_1$ and $\Gamma_2$ is poor and average
respectively, while $\Gamma_3$ and $\Gamma_4$ have very large
encodability. It was shown that the degree of encodability is mainly a
geometrical property of the structure and not too sensitive to the
number of amino acid classes or the values of interaction parameters
\cite{wing,steric,mprep}. For this reason we expect that the relative
encodability of $\Gamma_1-\Gamma_4$ remain different when using all
the three sets of parameters.

\section{Results and discussion}

\begin{figure}
\begin{center}
\epsfig{figure=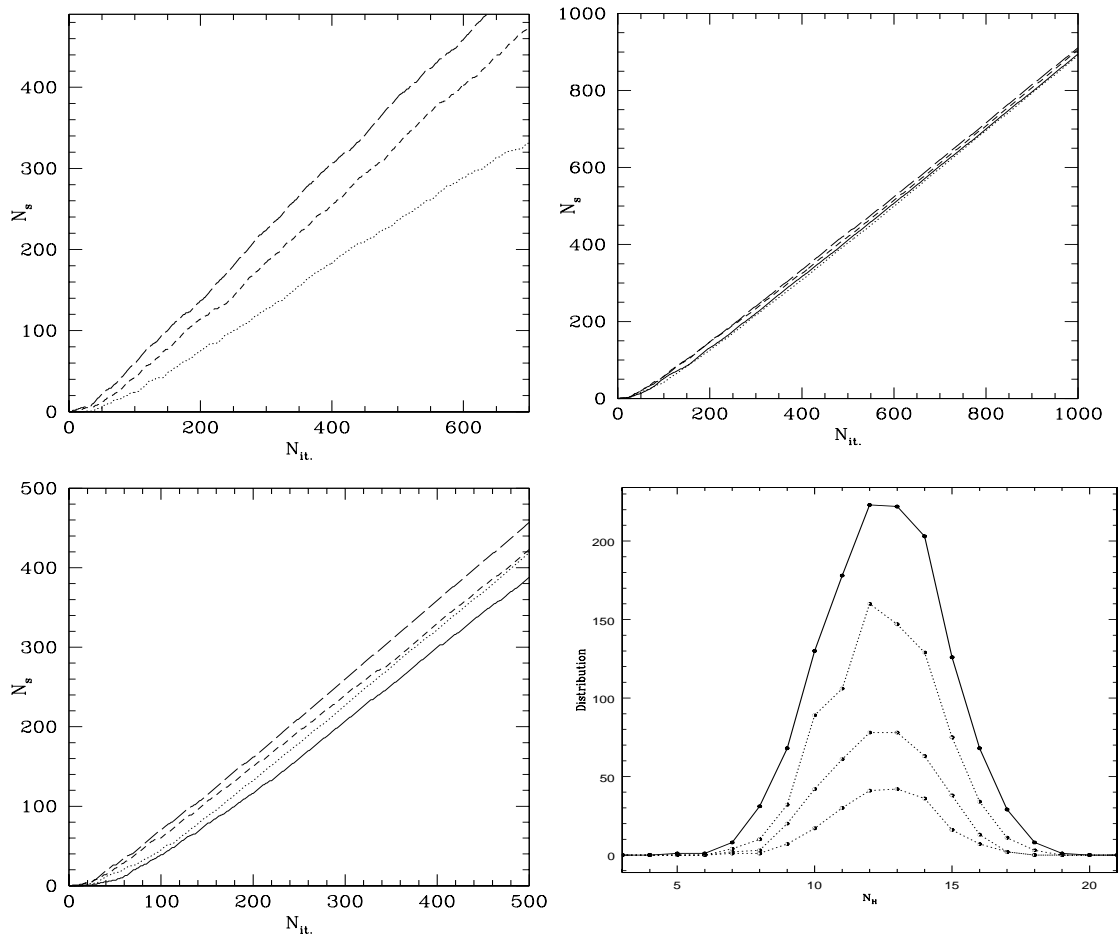,width=6.0in,height=5.0in}
\end{center}
\caption{Number of extracted solutions versus the number of iterations
for HP interactions (top left),  6 amino acid classes (top right) and 
 20 classes (left bottom). 
 The ideal curve, corresponding to efficiency 1, should have
slope 1. Plots referring to structures $\Gamma_{1}$, $\Gamma_2$,
$\Gamma_3$, $\Gamma_4$ are denoted with continuous, dotted, dashed and
long-dashed lines respectively. Bottom right panel  represents the histogram of
the number of extracted solutions a a function of sequence composition
(HP model). Curves pertain to an HP-design attempt on structure
$\Gamma_4$ at different values of $N_{it}: 200, 400, 800, 1719$. It
can be seen that the efficiency of the design technique is independent
of the sequence composition.}
\label{fig:perf}
\end{figure}

The ``dynamical'' performance of the algorithm can be seen in Figs.
\ref{fig:perf}a-c. The plots show the number of solutions retrieved as
a function of the number of iterations at a ``physiological''
temperature equal to $0.1,\ 10.0$ and $0.7$ for 2, 6 and 20 classes of
amino acids, respectively. The different values of the physiological
temperature are related to the different energy scales of the
interactions.

It can be seen that, after an initial
transient, the performance of the method (given by the slope of the
curves) is very high. In particular, for a large number of classes, it
is nearly equal to 1 for all structures. Table \ref{tab:nsol} provides
a quantitative summary of the performance of the method. For the HP
model, first column of Table \ref{tab:nsol}, the method was iterated until it
could not find further solutions with (estimated) folding temperature
greater that 0.1 . For the cases of 6 and 20 classes, a very large
number of solutions exist. Hence, we stopped the procedure after 1000
or 500 iterations, depending on the number of classes.

\begin{figure}
\centering
\epsfig{figure=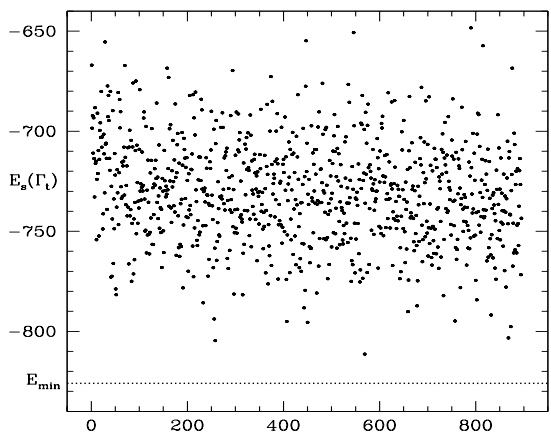,width=3.0in,height=2.5in}
\caption{Energy of the solutions found for structure
$\Gamma_4$ (6-class model) at fixed composition
$(4,4,4,5,5,5)$}.
\label{fig:en}
\end{figure}

\begin{figure}
\centering
\epsfig{figure=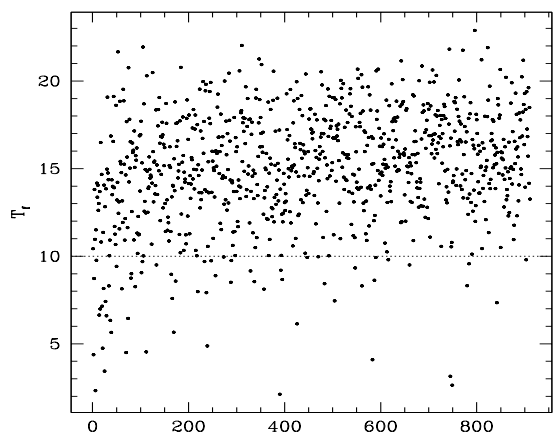,width=3.0in,height=2.5in}
\caption{Folding temperatures of solutions designing structure
$\Gamma_4$ (6-class model) as a function of the order of extraction.
Very few solutions turn out to have a folding temperature below the
simulation temperature $T=10$ (shown with a dotted line).}
\label{fig:tf}
\end{figure}

\begin{figure}
\centering
\epsfig{figure=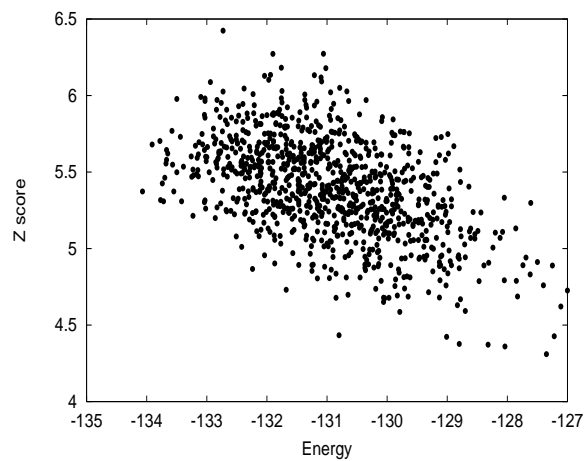,width=3.0in,height=2.5in}
\caption{
Scatter plot of the $Z_{score}$ against native-state energy of
extracted solutions designing structure $\Gamma_1$. The data are for a
20-letter alphabet of amino acids at fixed and nearly uniform composition.
}
\label{fig:zscore}
\end{figure}

\begin{figure}
\centering
\epsfig{figure=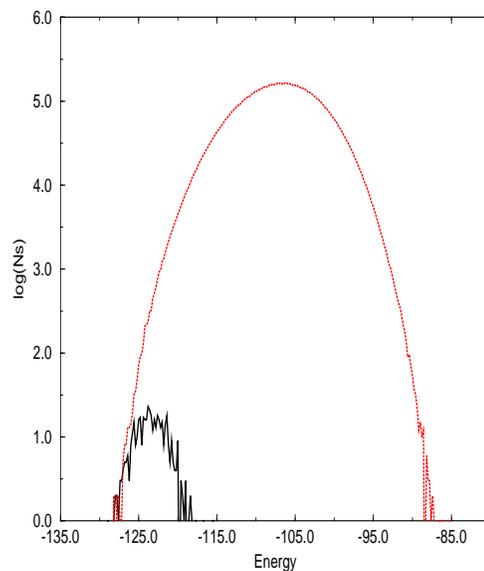,width=3.0in,height=3.5in,angle=270}
\caption{
Solid line: distribution (in arbitrary units) of solutions
(good sequences) to the design problem on structure $\Gamma_1$ (20 letter
alphabet). The dotted line denotes the distribution containing
 bad sequences. The data was obtained by randomly
sampling $10^7$ sequences with fixed composition.
}
\label{fig:rand_seq}
\end{figure}

An appealing feature is that the extracted solutions show no bias for
sequence composition (see Fig. \ref{fig:perf}d) or ground-state
energy. This can be seen in Fig. \ref{fig:en}, where we have plotted
the energies of 1000 designed solutions of fixed composition for the
6-classes case. Solutions do not exhibit packing around the minimum
energy ($\approx -830$) and their energy spread is fairly wide (the
estimated maximum energy is $\approx -170$). Furthermore, for each
extracted sequence we also calculated its folding temperature, to
compare it with $1/\beta$. As we remarked, if all the significant
competitors of $\Gamma_t$ were included in $D$, then sequences
satisfying (\ref{solution2}) should have folding temperatures greater
than $1/\beta$. As shown in the typical plot of Fig. \ref{fig:tf} this
is almost always the case, ensuring that solutions can be extracted
with a desired thermal stability. 
An alternative measure of the thermal stability connected to the
cooperativity and rapidity of the folding process is the
$Z_{score}$. For a sequence, $s$, designing structure $\Gamma$, the
$Z_{score}$ is defined as \cite{zscore}:
\begin{equation}
Z_{score}=\frac{\langle H_s\rangle-H_s(\Gamma)}{\sigma_s},
\end{equation}
where $\langle H_s\rangle$ is the average energy over the maximally compact 
conformations and $\sigma_s$ the standard deviation of the energy in this 
ensemble.
Fig. \ref{fig:zscore} shows a scatter plot of extracted solutions for target
structure $\Gamma_1$ for the 20-letter case. It can be seen that there
exist solutions with very high $Z_{score}$ throughout the displayed energy
range. This proves the usefulness of the novel design technique which
has no bias in native-state energy. In fact, it allows to collect
equally good folders with a wide range of native-state energy (and
hence very different sequences). This ought to be useful in realistic
design contexts, where among all putative design solutions one may
wish to retain those with specific amino acids in key protein sites.
The ability to select sequences across the whole energy range
highlights the efficiency of the technique.
In fact, as shown in Fig. \ref{fig:rand_seq}, away from the lowest energy edge, 
the fraction of good sequences over the total ones with the same energy is
minuscule (note the logarithmic scale).
Our method is able to span across the whole energy range without
restricting to those of minimal energy, which are a negligible
fraction of the total solutions.

\begin{figure}
\begin{center}
\epsfig{figure=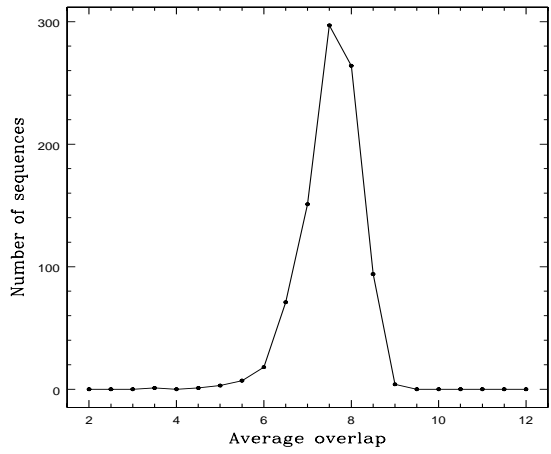,width=3.0in,height=2.5in}
\end{center}
\caption{
Histogram of the average overlap (sequence identity)
of solutions for $\Gamma_4$ (6-class model). 
For a given sequence the average overlap is calculated 
over all other extracted solutions.
}
\label{fig:seqov}
\end{figure}

Finally, we analysed the degree of
mutual similarity between extracted solutions. For the 6-classes case,
the sequence similarity between solutions was rather low, being around
20\%, as can be seen in Fig. \ref{fig:seqov}. This rules out the
possibility that solutions correspond to few point mutations of a
single prototype sequence.

One of the most significant features of the novel design procedure is
that the number of structures, $D$, used to calculate the approximate
free energy, (\ref{freeappr}), can be kept to a negligible fraction of
the total structures and yet allow a very efficient design. This is
proved even more strikingly by a further test of our design strategy
in the whole space of both compact and non-compact conformations. We
carried out a design of structure $\Gamma_2$ by using the HP
parameters with the constant $\alpha$ set to 0. This amounts to allow
for non-compact conformations to be native states. Since it is
unfeasible to explore this enlarged structure space, step 2 was
carried out with a stochastic Monte Carlo process, as described in
refs. [\onlinecite{11,M98a}], which generated dynamically growing
low-energy conformations at a suitable fictitious Monte Carlo
temperature. The correctness of the putative solutions was carried out
by using an algorithm known as Constrained Hydrophobic Core
Construction (CHCC)\cite{28,29}. The algorithm relies on an efficient
pruning of the complete search tree in finding possible low energy
conformations for a sequence. At the heart of the algorithm is the
observation that the most energetically convenient conformations for
the hydrophobic monomers is to form a compact, cubic-like, core. This
ideal situation may not be reachable for arbitrary sequences, due to
frustration effects; these are taken systematically into account to
build a compact core with a number of cavities sufficient to expose P
singlets (i.e. a P flanked by two H monomers in the sequence) on the
surface, which is energetically more effective than burying them in
the core. Then, exhaustive search algorithms are used to check the
compatibility of a sequence with cores of increasing surface area
(i.e. decreasing energy). A detailed description of the method can be
found in Refs. [\onlinecite{28,29}]. The time required by CHCC to find
the ground state energy of a sequence increases significantly, on
average, with the increase of the number of H residues. For this
reason we limited the search for design solutions to sequences with
$n_H=13$. The solutions, obtained in about one hundred iterations,
appears in table (\ref{C2-solutions}). All the 23 extracted solutions
had $\Gamma_2$ as the unique ground state among the compact
structures, and 17 of them retained $\Gamma_2$ as ground state even
when non-compact structures are considered. Given the vastity of the
enlarged structure space this represent a remarkable result.

\section{Conclusions}

We have presented a novel approach to protein design that encompasses
negative design features. Taking the latter into proper account has
been shown to be crucial for a succesful protein design. From a
practical point of view this amounts to calculating the conformational
free energy of all sequences which are candidate solutions. This
computational intensive task is kept to a minimum in our scheme thanks
to the identification of a limited number of structures which are
close competitors of the target conformations. The strategy, is easy
to implement and has been tested on minimalist models. The method
appears to be very efficient and reliable for a variety of different
sets of amino acid interactions. Contrary to other design techniques,
the extracted solutions show no bias in sequence composition or native
state energy and can be chosen to have a desired thermal stability.

We thank J. Banavar, G. Morra, F. Seno and G. Settanni for discussions
and suggestions. We acknowledge support from the Istituto Nazionale di
Fisica della Materia.

\newpage

\begin{table}
\begin{tabular}{|l|c|r|}\hline
&relative structures&Enc.\\\hline
$\Gamma_1$&\small URDDLLFFRBRFULLBBUFFRRBDLU&25\\\hline
$\Gamma_2$&\small UURFLFDBRBDFLFRRBBUUFFLDRB&337\\\hline
$\Gamma_3$&\small UURRFDLULDDFUURRDDBBULDFFU&1224\\\hline
$\Gamma_4$&\small UURRDLFFRBULLDDRBRFFLLUURR&1303\\\hline
\end{tabular}
\caption{The four structures used for benchmarking the design
strategy. The conformations are encoded in bond directions: U, up; D,
down; L, left; R, right; F, forward; B, backward. The encodability in
the rightmost column is defined as the number of sequences admitting
the corresponding structure as their unique native state (HP
interactions are assumed).}
\label{structures}
\end{table}

\begin{table}
\begin{tabular}{|r|r|r|r|r|r|r|}\hline
 &1 &2 &3 &4 &5 &6\\\hline
1&-50.00 &-20.49 &-38.20 &-6.65 &-43.65 &-10.63\\\hline
2&-20.49 &-14.91 &-18.13 &-4.00 &-15.56 & -3.81\\\hline
3&-38.20 &-18.13 &-35.75 &-5.07 &-23.96 &-26.02\\\hline
4& -6.65 & -4.00 &-5.07 & -1.65 & -5.17 & -9.47\\\hline
5&-43.65 &-15.56 &-23.96 &-5.17 &-43.71 &-18.63\\\hline
6&-10.63 & -3.81 &-26.02 &-9.47 &-18.63 &-26.70\\\hline
\end{tabular}
\label{tab:6-class}
\caption{Energy parameters for the 6-class model. Parameters
obey the segregation principle \protect{\cite{L95}}.}
\end{table}

\begin{table}
\begin{tabular}{|l||r|r||r|r||r|r||}\hline
&\multicolumn{2}{c||}{HP}&\multicolumn{2}{c||}{6
classes}&\multicolumn{2}{c||}{20 classes}\\\hline &$N_{it}$ &$N_{sol}$
&$N_{it}$ &$N_{sol}$ &$N_{it}$ &$N_{sol}$\\\hline
$\Gamma_1$&62&8&1000&895&500&388\\\hline
$\Gamma_2$&722&337&1000&891&500&419\\\hline
$\Gamma_3$&1898&1219&1000&906&500&423\\\hline
$\Gamma_4$&1719&1297&1000&911 &500&457\\\hline
\end{tabular}
\caption{Number of extracted solutions, $N_{sol}$, after $N_{it}$
iterations of the design procedure. For the HP model $N_{it}$ is the
number of iterations at which the iterative scheme stopped. It was
verified that the 1297 extracted solutions for structure $\Gamma_4$
have a folding temperature between 0.15 and 0.6.}
\label{tab:nsol}
\end{table}

\begin{table}
\begin{tabular}{c}\hline
Correct solutions \\ \hline

010111001110110001010100001 \\ 
000011011100111101000100101 \\ 
000010011100111101000101101 \\ 
000010000111100101110101101 \\ 
000010010100101111000110111 \\ 
000010000110100111010110111 \\ 
000010110100100111000101111 \\ 
000010000110100111010110111 \\ 
000010110100100111000101111 \\ 
000010010100101111000101111 \\ 
000010000110100111010101111 \\ 
000010110100100101000111111 \\ 
000010010100101101000111111 \\ 
000010100100100111000111111 \\ 
000010010100100111000111111 \\ 
000010000110100101010111111 \\ 
000010000100100111010111111 \\ 
\hline \hline
Incorrect solutions \\ \hline
 
110010001110110001010101001 \\ 
010011001110110100010100101 \\ 
110010001100110101000101101 \\ 
100011001100110101000101101 \\ 
000010100100100111010101111 \\ \hline \hline
\end{tabular}
\caption{Extracted solutions for structure $\Gamma_2$. The design
attempt was carried out in the whole space of conformations with
arbitrary degree of compactness.}
\label{C2-solutions}
\end{table}
%


\onecolumn

%
%
\newpage

\twocolumn


\end{document}